\documentclass[3p,twocolumn,number,sort&compress]{elsarticle}
\usepackage{pifont}
\usepackage{graphicx}
\usepackage[numbers]{natbib}
\usepackage{geometry}
\usepackage{amsmath}
\usepackage{amsfonts}
\usepackage{amssymb}
\usepackage{hyperref}
\usepackage[format=hang,labelfont=bf,textfont=small]{caption}
\usepackage{txfonts}
\usepackage{subfig}
\usepackage{dcolumn}
\usepackage{booktabs}
\begin{document}
\author[hgw]{C. Droese\corref{cor1}}
\author[heid]{K. Blaum}
\author[gsi]{M. Block}
\author[heid]{S. Eliseev}
\author[gsi]{F. Herfurth}
\author[mai]{E. Minaya Ramirez}
\author[rus,rus1]{Yu. N. Novikov}
\author[hgw]{L. Schweikhard}
\author[rus1]{V. M. Shabaev}
\author[rus1]{I. I. Tupitsyn}
\author[pol]{S. Wycech}
\author[dre]{K. Zuber}
\author[rus1]{N. A. Zubova}
\cortext[cor1]{Email corresponding author: c.droese@gsi.de; Phone: +496151712114; Fax: +496151713463; Current Address: Gesellschaft f\"ur Schwerionenforschung Planckstra\ss e 1 64291 Darmstadt Germany}
\address[hgw]{Institut f\"ur Physik, Ernst-Moritz-Arndt-Universit\"at, 17487 Greifswald, Germany}
\address[heid]{Max-Planck-Institut f\"ur Kernphysik, Saupfercheckweg 1, 69117 Heidelberg, Germany}
\address[gsi]{GSI Helmholtzzentrum f\"ur Schwerionenforschung, Planckstra{\ss}e 1, 64291 Darmstadt, Germany}
\address[mai]{Helmholtz-Institut Mainz, Johannes Gutenberg-Universit\"at, 55099 Mainz, Germany}
\address[rus]{Petersburg Nuclear Physics Institute, Gatchina, 188300 St. Petersburg, Russia}
\address[rus1]{St. Petersburg State University, 198504 St. Petersburg, Russia}
\address[pol]{Soltan Institute for Nuclear Studies, PL-00-681 Warsaw, Poland}
\address[dre]{Institut f\"ur Kern- und Teilchenphysik, Technische Universit\"at, 01069 Dresden, Germany}
\title{Probing the nuclide $^{180}$W for neutrinoless double-electron capture exploration}

\maketitle

\section*{Abstract}
The mass difference of the nuclides $^{180}$W and $^{180}$Hf has been measured with the Penning-trap mass spectrometer SHIPTRAP to investigate $^{180}$W as a possible candidate for the search for neutrinoless double-electron capture. The $Q_{\epsilon\epsilon}$-value was measured to 143.20(27)\,keV. This value in combination with the calculations of the atomic electron wave functions and other parameters results in a half-life of the $0^{+}\rightarrow 0^{+}$ ground-state to ground-state double-electron capture transition of approximately 5$\cdot 10^{27}$years/$<m_{\epsilon\epsilon}[eV]>^{2}$.  
\section{Introduction}
One of the open questions in elementary particle physics is whether neutrinos are Majorana or Dirac particles. The answer to this question could be given by the observation of neutrinoless double-electron transformations, \textit{i.e.} the double beta-particle emission, double-electron capture of orbital electrons, or a mixture of positron emission and electron capture.\\
Even though a resonant enhancement of neutrinoless double-electron capture was suggested long ago \cite{Win55,Ber83}, only nowadays the use of Penning traps allows an assessment of different candidate nuclides \cite{Bla06,Bla10}. Based on a very precise $Q_{\epsilon\epsilon}$-value measurement the resonant enhancement factor \cite{Ber83,EliGd11}
\begin{equation}
F= \frac{\Gamma_{2h}}{\Delta^{2}+\Gamma^{2}_{2h}/4} 
\end{equation}
can be determined to identify the most suitable candidates. In Eq. (1) $\Gamma_{2h}$ is the width of the double-electron hole \cite{Cam01} and $\Delta=Q_{\epsilon \epsilon}-B_{2h}-E$ is the degeneracy factor where $B_{2h}$ represents the binding energy of the captured electron pair and $E$ is the energy of the excited state in the daughter nuclide.\\
Several nuclides that had been listed as candidates \cite{Fre05} have already been probed.
As an example, a recent $Q_{\epsilon\epsilon}$ measurement \cite{Rah09} ruled out a resonant enhancement in $^{112}$Sn. An even more stringent negative result was obtained for another candidate, $^{74}$Se \cite{Kol10,Mou10}. Subsequent Penning-trap measurements of the $Q_{\epsilon\epsilon}$-values of $^{96}$Ru, $^{162}$Er, $^{168}$Yb, and $^{136}$Ce also excluded all of them from the list of suitable candidates for the experimental search for neutrinoless double-electron capture \cite{Eli11,Kol11}.\\
More promising candidates for neutrinoless double-electron capture are systems with ground-state to ground-state transitions ($0^{+}\rightarrow 0^{+}$) \cite{Bla10}. Their major advantage is a larger nuclear matrix element $M_{\epsilon\epsilon}$ and a larger phase space compared to the ground-state to excited-state transitions. The maximum capture rate is expected if the resonant enhancement criterion is fulfilled for the capture of two K-shell electrons in heavy nuclides. 
Two candidates for such types of transitions, proposed in \cite{Bla10}, have recently been explored with the Penning trap system SHIPTRAP: $^{152}$Gd$\rightarrow ^{152}$Sm \cite{EliGd11} and $^{164}$Er$\rightarrow ^{164}$Dy  \cite{EliDy11}. The estimated half-life for $^{152}$Gd of $10^{26}$ years/$<m_{\epsilon\epsilon}[eV]>^{2}$ ($<m_{\epsilon\epsilon}[eV]>$ is the effective Majorana neutrino mass given in electron volts) assesses this nuclide as the best candidate known to date for the search for neutrinoless double-electron capture.\\
The present study explores the $0^{+}\rightarrow 0^{+}$ transition of $^{180}$W to $^{180}$Hf. This isotope remains the last ground-state to ground-state transition candidate that has not been investigated before our studies. The mass difference $Q_{\epsilon\epsilon}$ given in the Atomic-Mass Evaluation 2003 is 144.4(45)\,keV \cite{Aud03}, which is not accurate enough to evaluate the resonant enhancement factor.
\section{Experimental Setup}
\begin{figure}[!htb]
\centering%
\includegraphics[scale=0.4]{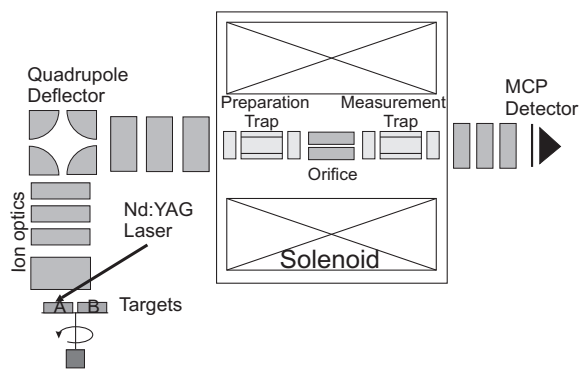}%
\caption[Excitation energies of Rn]{A sketch of the SHIPTRAP setup. For details see text.}%
\label{fig:Q-value_SetUp4}%
\end{figure}

\begin{figure}[!htb]
\centering%
\includegraphics[scale=0.3]{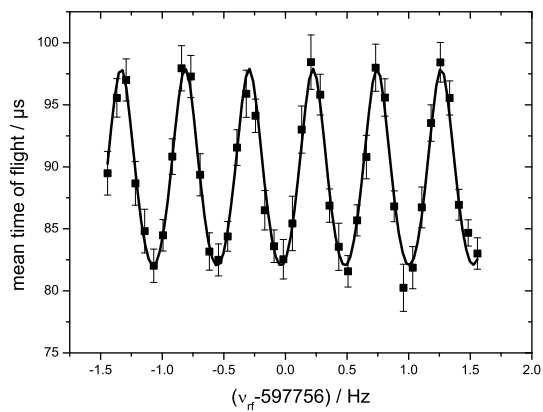}%
\caption[Excitation energies of Rn]{Time-of-flight ion-cyclotron-resonance of $^{180}$W$^{+}$ for a Ramsey-excitation scheme with two excitation pulses of 100\,ms separated by a waiting time of 1800\,ms. The solid line is a fit of the expected line shape \cite{Kre07} to the data.}%
\label{fig:W_Res}%
\end{figure}

The $Q_{\epsilon\epsilon}$-value of $^{180}$W reported here was measured by high-precision Penning trap mass spectrometry with SHIPTRAP \cite{Blo07}. The part of the SHIPTRAP apparatus used for the present measurements is sketched in figure \ref{fig:Q-value_SetUp4}.
All ions were produced via laser ablation from enriched samples ($^{180}$W metal with 89.1\% and $^{180}$HfO with 93.89\% isotopic enrichment) on a stainless steel backing \cite{Cha07}. The ions were guided with electrostatic lenses and a quadrupole deflector into a tandem Penning trap placed in a 7-T superconducting solenoid.\\
In the preparation trap possible contaminants were removed by mass-selective buffer-gas cooling \cite{Sav91} from the ion ensemble before it was transferred to the measurement trap, where the cyclotron-frequency measurement of the ion of interest was performed.
To this end the time-of-flight ion-cyclotron-resonance (ToF-ICR) technique \cite{Gra80} was utilized to determine the cyclotron frequency $\nu_{c}=qB/(2\pi m)$ with the magnetic-field strength $B$ and the ion's charge-to-mass ratio $q/m$. The charged particles were detected by a micro-channel-plate (MCP) detector. A Ramsey-excitation scheme was applied \cite{Geo07,Geor07} using an excitation pattern of two radiofrequency pulses with a duration of 50\,ms or 100\,ms, separated by a waiting time of 900\,ms or 1800\,ms, respectively. An example of such a ToF-ICR resonance of $^{180}$W$^{+}$ is shown in figure \ref{fig:W_Res}.\\
To determine which is the central minimum of the Ramsey curve, the cyclotron frequency was checked regularly by a single-pulse ToF-ICR measurement with an uncertainty of about 50\,mHz.\\
The resonances of $^{180}$W$^{+}$ and $^{180}$Hf$^{+}$ were taken alternately. For the frequency-ratio determination the cyclotron frequency of one nuclide was linearly interpolated to the time of the measurement of the other one. Thus, due to active stabilization systems \cite{Dro11} any residual magnetic-field fluctuations during the short time between the resonance measurements were negligible. Due to the extremely small mass difference of the two species and the identical ion production as well as preparation mechanisms systematic uncertainties can be neglected on the present level of statistical uncertainties \cite{Rou11}. Note that no absolute mass values are deduced from the present measurements, but only the mass difference between the two close-lying ion species. 
For the analysis only those events with an ion number of less or equal to 5 ions per cycle were considered in order to exclude frequency shifts due to ion-ion interactions \cite{Bol90}. This data was split up into five different subsets according to the number of detected ions per measurement cycle. For each subset a frequency ratio was determined \cite{Kel03}. Those ratios that showed an unreasonably high scattering for different numbers of ions were removed from the measurement.\\ 
Further details of the data analysis of $Q_{\epsilon \epsilon}$-value measurements at SHIPTRAP will be given in a forthcoming publication \cite{Rou11}.
\section{Results} 
\begin{figure}[!h]
\centering%
\includegraphics[scale=0.3]{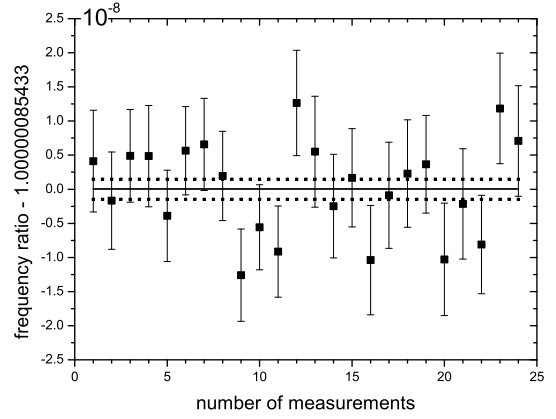}%
\caption[Excitation energies of Rn]{Frequency ratios $\nu_{c}(^{180}$Hf$^{+}$/$^{180}$W$^{+}$) (points) with the average value (solid line) and the one sigma statistical uncertainty (dotted lines).}%
\label{fig:Ratios_vs_num_W}%
\end{figure}

\begin{figure*}[!htb]
\centering%
\includegraphics[scale=1]{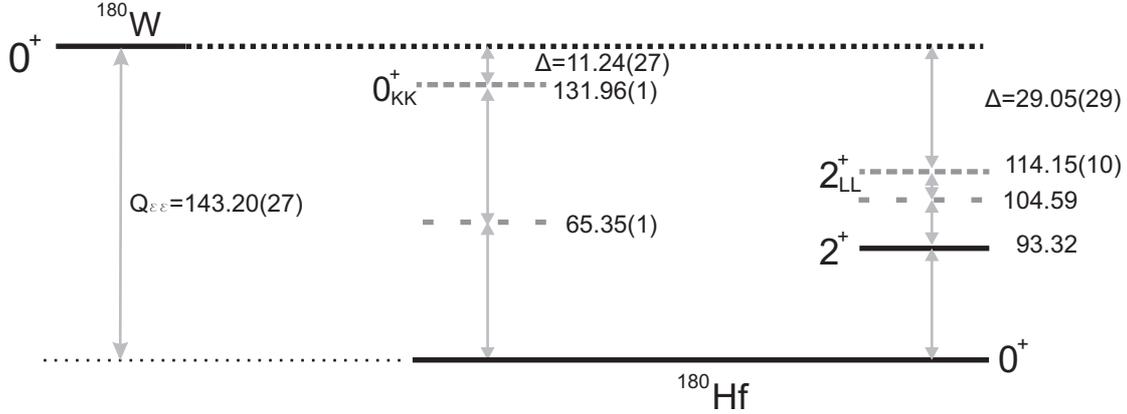}%
\caption[Excitation energies of Rn]{Partial level scheme of the double-electron capture with the $Q_{\epsilon\epsilon}$-value including the energy difference $\Delta$ to the ground-state with the double K-shell electron binding energy 0$^{+}_{KK}$ and the less favorable excited-state transition to the atomic shell orbitals L$_{1}$ and L$_{3}$ 2$^{+}_{LL}$ of $^{180}$Hf in keV.}%
\label{fig:level scheme}
\end{figure*}

The resulting frequency ratios $\nu_{c}(^{180}$Hf$^{+})$/$\nu_{c}(^{180}$W$^{+})$ are shown in figure \ref{fig:Ratios_vs_num_W}. The error bars of the individual data points give the statistical uncertainties. The mean frequency ratio is
\begin{equation}
\nu_{c}(^{180}\text{Hf}^{+})/\nu_{c}(^{180}\text{W}^{+})=1+8.5433(165)\cdot 10^{-7}.
\end{equation}
The $Q_{\epsilon\epsilon}$-value of the decay is given by
\begin{eqnarray}
Q_{\epsilon\epsilon} & = & m(^{180}\text{W})-m(^{180}\text{Hf}) \nonumber\\
 & = & [m(^{180}\text{Hf})-m_{e}]\cdot\left[ \frac{\nu_{c}(^{180}\text{Hf}^{+})}{\nu_{c}(^{180}\text{W}^{+})}-1\right],
\end{eqnarray}\\
where $m$($^{180}$Hf) is the atomic mass of the daughter nuclide and $m_{e}$ is the electron mass.\\
The resulting energy difference between the initial and final species of the double-electron capture transition $^{180}$W$\rightarrow ^{180}$Hf is $Q_{\epsilon\epsilon}$=143.20(27)\,keV. The uncertainty of the $Q_{\epsilon\epsilon}$-value is reduced by a factor of 17 in comparison to the old value of $Q_{\epsilon\epsilon}$=144.4(45)\,keV deduced from the mass values given in \cite{Aud03}. The value 131.96(1)\,keV for the K-shell double-electron hole binding energy $B_{2h}$ was taken from \cite{Kri10}. The degeneracy factor is calculated to be $\Delta$=11.24(27)\,keV. The width of the double-electron hole $\Gamma_{2h}$=71.8\,eV has been deduced from \cite{Cam01}. Thus, the resonance condition is not fulfilled.\\
Energetically less favorable is the capture to a nuclear excited state in $^{180}$Hf with an energy of 93.3243(20)\,keV \cite{Wu03} and a spin parity of $I$=2$^{+}$. The difference between the mother and daughter states is 29.05(27)\,keV for the atomic shell orbitals L$_{1}$ and L$_{3}$ and their binding energies, 11.27\,keV and 9.56\,keV, respectively.
Based on all these parameters the transition to the excited nuclear state is unfavorable. A partial level scheme of the daughter nuclide $^{180}$Hf is shown in figure \ref{fig:level scheme}.\\
The capture probability was deduced from \cite{Ber83} and is given by
\begin{equation}
\Lambda_{0\nu\epsilon\epsilon}=\frac{2g_{A}^{4}G_{F}^{4}(\text{cos}\theta_{c})^{4}}{(4\pi R)^{2}}<m_{\epsilon\epsilon}>^{2}\vert M_{\epsilon\epsilon}\vert^{2}P_{\epsilon\epsilon}F.
\end{equation}
In order to estimate $\Lambda_{0\nu\epsilon\epsilon}$ the product of the electron wave functions $P_{\epsilon\epsilon}=\vert\Psi_{h1}\vert^{2}\vert\Psi_{h2}\vert^{2}$ for two captured electrons $h_{1}$ and $h_{2}$ in the nucleus was determined using the one-configuration Dirac-Fock method for an extended nucleus \cite{Bra77}. The one-electron densities of the K-shell were calculated for the ground state of neutral $^{180}$W at the center and the root-mean-square radius of the atomic nucleus. Then, the mean value of these densities was used in the calculation of $P_{\epsilon\epsilon}$. The characteristic value $\vert M_{\epsilon\epsilon}\vert^{2}$=9 for transitions between heavy nuclides was used \cite{Kri10}. In Eq. (4) $g_{A}$ stands for the axial-vector nucleon coupling constant, $G_{F}$ is the weak coupling constant, $\theta_{c}$ is the Cabibbo angle and $R$ is the nuclear radius.
As compared to other candidates for neutrinoless double-electron capture the absence of the energy degeneracy is partly compensated by the high values of $P_{\epsilon\epsilon}$ of 3.2$\cdot 10^{11}$ and the nuclear matrix element. We estimate the half-life of $^{180}$W to
\begin{equation}
T_{1/2}=\frac{ln2}{\Lambda_{0\nu\epsilon\epsilon}}\approx\frac{5\cdot 10^{27}}{\langle m_{\epsilon\epsilon}[eV]\rangle^{2}}\text{years.}
\end{equation}
In \cite{Suj04} a radiative neutrinoless double-electron capture was predicted, particular for the case of $^{180}$W. In a simplified description one of the electrons captured from the initial atomic state emits a photon during that capture process. The energy of the emitted photon is equal to the degeneracy factor $\Delta_{KL}$, in which $B_{2h}$ is the binding energy of captured K- and L-shell electrons. The transition probability of the process increases if the energy of the photon is equal to the energy difference of the atomic states $1S-2P$. In this particular case, a resonant enhancement of double-electron capture occurs. Thus, the resonance condition for radiative neutrinoless double-electron capture appears when $\Delta_{KL}=E(1S-2P)$, whereas the resonance condition for the nonradiative transition analysed in this work is fulfilled when $\Delta_{KK}$ is approaching the value zero.
The transition $^{180}$W$\rightarrow ^{180}$Hf was considered with the new measured value of $Q_{\epsilon\epsilon}$=143.20(27)\,keV. With the formulas presented in \cite{Suj04} we estimated a half-life of 4$\cdot 10^{27}$ years/$<m_{\epsilon\epsilon}[eV]>^{2}$.\\
\section{Summary and Conclusion}
This work continued the search for a suitable candidate for a future experiment dedicated to the observation of neutrinoless double-electron capture. The focus of this work was on the investigation of the ground-state to ground-state double-electron capture $0^{+}\rightarrow 0^{+}$ in $^{180}$W. This transition has a rather large nuclear matrix element, and for the case of two K-shell electron captures has a high overlap of the electron wave functions. Except for the investigated nuclide only a very limited number of possible neutrinoless double-electron capture candidates, \textit{i.e.} $^{152}$Gd and $^{164}$Er, fulfil these requirements.
Here we measured the atomic mass differences of $^{180}$W and its daughter nuclide $^{180}$Hf resulting in a value of $Q_{\epsilon \epsilon}$=143.20(27)\,keV. Although the obtained value does not result in an energy degeneracy of the mother and daughter states, a rather high neutrinoless double-electron capture probability with a half-life of 5$\cdot 10^{27}$ years/$<m_{\epsilon\epsilon}[eV]>^{2}$ is obtained. 
For the process of radiative neutrinoless double-electron capture \cite{Suj04} a half-life close to this value of 4$\cdot$10$^{27}$years/$<m_{\epsilon\epsilon}[eV]>^{2}$ was estimated.
Both values lie between the results of the other two known ground-state to ground-state transition candidates, which both had already been investigated, $^{152}$Gd with a half-life of about $10^{26}$years/$<m_{\epsilon\epsilon}[eV]>^{2}$ and $^{164}$Er with 1$\cdot 10^{30}$years/$<m_{\epsilon\epsilon}[eV]>^{2}$.
The theoretical prediction from \cite{Kri10} combined with the experimental results and the matrix element given above yields a half-live of 1.2$\cdot 10^{28}$years/$<m_{\epsilon\epsilon}[eV]>^{2}$.
\section*{Acknowledgments}
We acknowledge the support of the German BMBF (Grants 06GF186I, 06GF9103I, 06DD9054) and by the WTZ Grant RUS-07/015. E.M. thanks the Helmholtz-Institute Mainz. Yu. N. thanks EMMI for his guest professorship. Financial support by the Max-Planck Society is acknowledged. The work of V. M. S., I. I. T. and N. A. Z. was supported by RFBR (Grant No. 10-02-00450).
\\

\end{document}